\documentclass[acmtosem,hyperref]{acmtrans2m} 

\usepackage[T1]{fontenc} 

\usepackage{charter}

\usepackage{listings}             

\usepackage{algorithmic} 

\usepackage{graphicx}

\usepackage{tabularx}
\newcolumntype{Y}{>{\centering\arraybackslash}X}

\usepackage{amsmath}

\acmVolume{}
\acmNumber{}
\acmYear{2011}
\acmMonth{Nov}

\usepackage{pdfpages}

\pdfoutput=1
\usepackage[colorlinks,linkcolor=blue,citecolor=blue,urlcolor=blue]{hyperref}

\title{Detecting Missing Method Calls As Violations of the Majority Rule}

\author{
Martin Monperrus, Mira Mezini\\
Technische Universit\"at Darmstadt\footnote{Martin Monperrus is now with the University of Lille and INRIA.}}
  
\begin{abstract}
When using object-oriented frameworks it is easy to overlook certain important method calls that are required at particular places in code. In this paper, we provide a comprehensive set of empirical facts on this problem, starting from traces of missing method calls in a bug repository. We propose a new system that searches for missing method calls in software based on the other method calls that are observable. Our key insight is that the voting theory concept of majority rule holds for method calls: a call is likely to be missing if there is a majority of similar pieces of code where this call is present. The evaluation shows that the system predictions go further missing method calls and often reveal different kinds of code smells (e.g. violations of API best practices).
\end{abstract}

\category{D 2.3}{Software Engineering}{Coding Tools and Techniques}

\newcommand{\code}[1]{{\ttfamily #1}}

\newcommand{\toolname}{DMMC}

\newtheorem{definition}{Definition}

\begin{document} 

\maketitle

\section{Introduction}

\emph{``Thanks for letting me know about [...] the missing method call''.} 
This was written by a programmer on an Internet forum\footnote{\scriptsize\url{http://www.velocityreviews.com/forums/t111943-customvalidator-for-checkboxes.html}}.
This quote indicates that missing method calls may be the source of software defects that are not easy to detect without assistance. 
Actually, problems related to missing method calls pop up in forums$^1$, in newsgroups\footnote{\scriptsize\url{http://dev.eclipse.org/mhonarc/newsLists/news.eclipse.tools/msg46455.html}}, in bug reports\footnote{\scriptsize\url{https://bugs.eclipse.org/bugs/show_bug.cgi?id=222305}}, in commit  texts\footnote{\scriptsize\url{http://dev.eclipse.org/viewcvs/index.cgi/org.eclipse.team.core/src/org/eclipse/team/core/mapping/provider/MergeContext.java?view=log}}, and in source code\footnote{\scriptsize\url{http://dev.eclipse.org/viewcvs/index.cgi/equinox-incubator/security/org.eclipse.equinox.security.junit/src/org/eclipse/equinox/security/junit/KeyStoreProxyTest.java?view=co}}. 
For a more systematic analysis of the problem, we performed a comprehensive study in a well-delimited scope: the Eclipse Bug Repository contains at least 115 bug reports related to missing method calls (cf. section \ref{mcm-in-eclipse-bug-repository}).
The analysis shows that issues caused by missing method calls are manifold\footnote{These issues are further discussed later in sections \ref{characterization} and \ref{case-studies}.}: they can produce obscure runtime exceptions at development time, they can be responsible of defects in limit cases, and they generally reveal code smells. 
These observations have motivated the work presented in this paper.

Our intuition is that missing method calls are a kind of \emph{deviant code}. Previous research proposed different characterizations of \emph{deviant code}. 
Engler et al. \cite{Engler2001} and Li et al. \cite{Li2005} proposed two different characterizations for procedural system-level code. Livshits et al. \cite{Livshits2005} characterized deviant code as instance of error patterns highlighted by software revisions. Wasylkowski et al. \cite{Wasylkowski2007} described an approach based on mining usage patterns and their violations. 
However, as we will further elaborate in section \ref{related}, the aforementioned proposals either are  not dedicated to object-oriented code and subsequently to missing method calls, 
or suffer from precision or scalability issues.

This paper presents  \emph{a new characterization of deviant code} suitable to detect missing method calls. The pieces of code that we consider are \emph{type-usages}. 
A type-usage is a list of method calls on a variable of a given type occurring somewhere within the context of a particular method body. 
In voting theory, the majority rule is the concept underlying most democratic electoral methods [p .3]\cite{Coughlin1992}, stating that what the majority chooses is the best.
Our intuition is that the concept of majority rule holds for type-usages: a type-usage is likely to host defects if there are few similar pieces of code and a majority of slightly different pieces of code.

We design a new metric called \emph{S-score}, which applies the concept of majority rule to type-usages of object-oriented programs.
Eventually, the \emph{S-score} measures the degree of deviance of type-usages: the higher the \emph{S-score}, the higher the type-usage smells.
Indeed, our approach produces warnings for type-usages whose \emph{S-score} is high.
The implementation of the approach is a {\bf D}etector of {\bf M}issing {\bf M}ethod {\bf C}all, which grounds its acronymic name: \toolname.

We use different techniques to evaluate the proposed approach. First, statistical methods are used to show that our characterization of deviant code makes sense for detecting missing method calls. Second, we propose and perform a quantitative evaluation based on the simulation of defects by degrading real software. The advantage of this evaluation technique is that it can be fully automated on a large scale while still involving likely defects. 
Our last evaluation technique uses our tool for revealing problems related to missing method calls in real software: the user-interface part of the Eclipse IDE codebase.
We present a set of case-studies that help to understand the value and the meaning of missing method calls predicted by our approach.

\vspace{2mm}
\noindent
To summarize, the contributions of this paper are:

\begin{itemize}

  \item A comprehensive set of empirical facts on the problems caused by missing method calls. We present +30 examples of real software artifacts affected by missing method calls, a comprehensive study of this problem in the Eclipse Bug Repository, and an extensive analysis of the missing calls that our tool found in Eclipse.

  \item A new characterization of \emph{deviant code}. This new characterization is built on observed method calls in real software. We also propose a technique to transform deviance data as concrete recommendations: the system tells the developer what methods seem to be missing at a particular place in code.
  
  \item A new strategy to evaluate code warning tools, based on the simulation of defects by degrading real software.

\end{itemize}

This article is a follow-up of a paper published at ECOOP'2010 \cite{Monperrus2010a}. 
It deepens the evaluation with new datasets (5 datasets while there was only one considered in \cite{Monperrus2010a}). The new experiments show that the results hold in different contexts, hence improve the external validity (the generalizability) of our evaluation. Also, when replicating the experiments, we obtained new insights into our evaluation process and improved it (further discussed in \ref{evaluation}), leading to more accurate results.

The reminder of the paper is structured as follows. 
In section \ref{characterization}, we elaborate on empirical facts about missing method calls.
Section \ref{contribution} presents our approach and the underlying algorithm.
Section \ref{evaluation} presents a quantitative evaluation of our system.
Section \ref{case-studies} contains qualitative case-studies, and shows that our system goes further missing method calls.
Related work is discussed in section \ref{related}.
Section \ref{conclusion} concludes the paper and sketches areas of future work.

\section{The Importance of Detecting Missing Method Calls}
\label{characterization}

This section presents empirical facts supporting the following claims: (a) problems related to missing method calls do happen in practice and can be difficult to understand, and (b) they are checked-in to the source code repository.

\subsection{Problems Related to Missing Calls are Real and Hard to Understand}
\label{sandra}

Let us tell a little story that shows that missing method calls are likely and can be the source of real problems. The story is inspired from several real world posts to Internet forums and mailing lists\footnote{e.g. \scriptsize\url{http://dev.eclipse.org/mhonarc/newsLists/news.eclipse.tools/msg46455.html} and \scriptsize\url{http://dev.eclipse.org/newslists/news.eclipse.platform.rcp/msg10075.html}}.
Sandra is a developer who wants to create a dialog page in Eclipse. She finds a class corresponding to her needs in the API named \code{DialogPage}. Using the new-class-wizard of Eclipse, she automatically gets a code snippet containing the methods to override, shown below:

\begin{verbatim}
public class MyPage extends DialogPage {
 @Override
 public void createControl(Composite parent) {
  // TODO Auto-generated method stub
 }
}
\end{verbatim}

Since the API documentation of \code{DialogPage} does not mention special things to do, Sandra writes the code for creating a control, a \code{Composite}, containing all the widgets of her own page. 
Sandra knows that to register a new widget on the UI, one passes the parent as parameter to the Composite constructor.
\begin{verbatim}
 public void createControl(Composite parent) {
   Composite mycomp = new Composite(parent);
   ....
 }
\end{verbatim}

Sandra get the following error message at the first run of her code (the error log is unfortunately empty)! 

\begin{verbatim}
An error has occurred. See error log for more details.
org.eclipse.core.runtime.AssertionFailedException
null argument:
\end{verbatim}

When extending a framework class, there are often some contracts of the form "\emph{call method x when you override method y}", which need to be followed. The Eclipse JFace user-interface framework expects that an application class extending \code{DialogPage} calls the method \code{setControl} within the method that overrides the framework method \code{createControl}.
However, the documentation of \code{DialogPage} does not mention this implicit contract; Sandra thought that registering the new composite with the parent is sufficient.

The described scenario pops up regularly in the Eclipse newsgroup\footnote{cf. the Google results of ``\url{setcontrol+site:http://dev.eclipse.org/mhonarc/newsLists/}'')} and shows that one can easily fail to make important method calls.
Furthermore, the resulting runtime error that Sandra got is really cryptic and it may take time to understand and solve it. 

Sandra had to ask a question on a mailing list to discover that this problem comes from a missing call to \code{this.setControl}. After the addition of \code{this.set\-Con\-trol(mycomp)} at the end of her code, Sandra could finally run the code and commit it to the repository; yet, she lost 2 hours in solving this bug related to a missing method call.

\subsection{Missing Method Calls Are Checked-in to Repositories}
\label{mcm-in-eclipse-bug-repository}
Missing method calls that are not all detected before leaving the developer's machine are actually committed code to the source code repository. To support this claim, we have searched for bug descriptions related to missing method calls in the Eclipse Bug Repository\footnote{\scriptsize\url{http://bugs.eclipse.org}}.

Our search process went through the following steps: 1) establish a list of syntactic patterns which could indicate a missing method call, 2) for each pattern of the list created in the previous step, query the bug repository for bug descriptions matching the pattern 3) read the complete description of each resulting bug report to assess whether it is really related to missing method calls.

To know that a report is really due to a missing method call or not, we read the whole sentence or paragraph containing the occurrence of the syntactic pattern. This gives a clear hint to assess whether this report is really related to a missing method call. For instance, bug \#186962 states that \emph{``setFocus in ViewPart is not called systematically''}: it is validated as related to missing method call; bug \#13478 mentions that \emph{``CVS perspective should be called CVS Repository Exploring''}: it is not related to our concern.

Table \ref{empirical-motivation} summarizes the results. For illustration consider the numbers in the first raw, which tell that 49 bug reports contain the syntactic pattern ``should call'', and 26 of them are actually related to missing method calls. In the 211 bug reports found by our syntactic patterns, 117 of them are actually related to missing method calls. This number shows that missing method call survive development time, especially if we consider that the number is probably an underestimation, since we may have missed other syntactic patterns. Indeed, we will also show in the evaluation section that we are able to find other missing method calls in Eclipse.

\begin{table}
\caption{The number of bug reports in the Eclipse Bug Repository per syntactic pattern related to missing method calls. The second column shows the number of occurrences of the pattern, the third one is the number of bug reports that are actually related to missing method calls after manual inspection.}
\label{empirical-motivation}
\begin{tabularx}{\textwidth}{|X|X|X|}
\hline
Pattern & ~Matched~~ & Confirmed \\
\hline
``should call'' & 49 & 26 (53\%)\\
``does not call'' & 39 & 28 (72\%)\\
``is not called'' & 36 & 26 (72\%)\\
``should be called''~~~~& 34 & 9 (26\%)\\
``doesn't call'' & 16 & 13 (81\%)\\
``do not call'' & 10 & 6 (60\%)\\
``are not called'' & 7 & 0 (0\%)\\
``must call'' & 7 & 4 (57\%)\\
``don't call'' & 6 & 2 (33\%)\\
``missing call''~~~~& 6 & 2 (33\%)\\
``missing method call''~~~~& 1 & 1 (100\%)\\
\hline
Total & 211& 117 (55\%)\\
\hline
\end{tabularx}
\end{table}

\subsection{Recapitulation}

These empirical facts show that a detector of missing method calls: (a) can help programmers like Sandra write better code in a shorter time, and (b) can help maintainers solve and fix bugs related to missing method calls. Also, from a quality assurance perspective, such a code warning tool lists places in code that are likely to contain missing method calls and that are worth being investigated before they produce a real bug or hinder maintenance.

\section{From Type-usages to Missing Method Calls}
\label{contribution}

This section presents an approach to qualify missing method calls as violations of the majority rule in object-oriented software. 

The historical rationale behind this approach is that, in our previous work, we came to the point that we should have a radically new viewpoint over code to lower the curse of high false positive rate (noted by Kim and Ernst in \cite{Kim2007}). Especially, we assumed that we need a move abstract viewpoint over code compared to related work (for instance abstracting over call ordering and control flow). 
Our proposal of abstraction over code is called \emph{type-usage} and is presented in section \ref{type-usages} below.  
and section \ref{relations-for-type-usages} introduces two relations between type-usages.
Then, section \ref{def:sscore} leverages those relations to define a measure of strangeness for type-usages.
Finally, section \ref{predicting-the-missing} uses all these concepts in an algorithm that predicts missing method calls.

\subsection{Type-Usage}
\label{type-usages}
Our approach is grounded on the concept of \emph{type-usage}, that can be defined as follows:

\begin{definition}
A type-usage is a list of method calls on a variable of a given type occurring in the body of a particular method. 
\end{definition}

Figure \ref{fig:typeusages} shows a code excerpt to illustrate this definition. 
In a method \code{create\-But\-ton}, there is one type-usage of type \code{Button}, which contains three method calls \code{Button.<init>, Button.setText(), Button.setColor()}.
There is exactly one type-usage per variable $x$ in source code, and a type-usage can be completely expressed by the following descriptors:
\begin{itemize}
\item $T(x)$ is the type of the variable containing the type-usage. If there are two variables of the same type in the scope of a method, they are two type-usages extracted.
\item $C(x)$ is the context of \code{x}, which we define as the signature of the containing method (i.e. name, and ordered parameter types in Java)
\item $M(x)$ is the set of methods invoked on \code{x} within $C(x)$. \end{itemize}

\begin{figure}
\caption{Extraction Process of Type-Usages in Object-Oriented Software.}
\label{fig:typeusages}
    \begin{minipage}{4cm}
    \begin{lstlisting}[language=java]
    class A extends Page {
      Button b;
      
      Button createButton() {
        b = new Button();
        b.setText("hello");
        b.setColor(GREEN);
        ...(other code)
        Text t = new Text();
        return b;
      }
    }
    \end{lstlisting}
    \end{minipage}\hfill \begin{minipage}{6cm}
    \begin{verbatim}
    T(b) = 'Button'
    C(b) = 'Page.createButton()'
    M(b) = {<init>, setText, setColor}

    T(t) = 'Text'
    C(t) = 'Page.createButton()'
    M(t) = {<init>}

    \end{verbatim}
    \end{minipage}
\end{figure}

Figure \ref{fig:typeusages} illustrates the conversion of Java code to type-usages. A code snippet is shown on the left-hand side of the figure;  the corresponding extracted type-usages are shown on the right-hand side of the figure. 
There are two extracted type-usages, for Button \code{b} and for Text \code{t}. The context is the method \code{createButton} for both. The set of invoked methods on \code{b} is $M(b)=\{<init>, setText, setColor\}$, \code{t} is just instantiated ($M(t)=\{<init>\}$).

The main insight behind our definition of type-usage is that it is a strong abstraction over code. In particular, it abstracts over call ordering and control flow as opposed to for example Anmons et al.'s ``scenario'' \cite{Ammons2002} or Nguyen et al.'s ``groums'' \cite{Nguyen2009}. For instance, let us consider that in the majority of cases, one has a call to \code{setText} before \code{setColor}. This does not mean that the opposite would be an anomaly, a deviation to correct usage. Our definition removes a sufficient amount of application-specific details and focuses on the core of using an API class: calling methods. That said, we don't claim that all bugs related to method calls can be caught by this abstraction. In particular, bugs related to object protocols are not addressed in our approach.

\subsection{Binary Relations between Type-Usages}
\label{relations-for-type-usages}

Let us now informally define two binary relations between type-usages: \emph{exact-similarity} and \emph{almost-similarity}. 

A type-usage is \emph{exactly-similar} to another type-usage if it has the same type, and is used in the method body of a similar method containing the same method calls. 
For instance, in Figure \ref{typeusage-example} the type-usage in class B (top-right snippet) is exactly-similar to the type-usage of class A (top-left snippet): (a) they both occur in the body of the method \code{Button createButton()}, i.e. they are used in the same context (the notion of ``context'' is defined in \ref{type-usages} as the signature of the containing method), and (b) they both have the same set of method calls. We use the term "similar" to highlight that at a certain level of detail the type-usages related by exact-similarity are different: variables names may be different, interlaced and surrounding code as well. 

A type-usage is \emph{almost-similar} to another type-usage if it has the same type, is used in a similar context and contains the same method calls plus another one. In figure \ref{typeusage-example} the type-usage in class C (bottom snippet) is almost-similar to the type-usage of class A (top-left snippet): they are used in the same context, but the type-usage in class C contains all methods of A plus another one: \code{setLink}. We need the term \emph{almost-similar} to denote that the relationship between two type-usages is more similar than different, i.e., there is some similarity, while being not \emph{exactly-similar}.

\begin{figure}
\caption{Examples of Exactly-Similar and Almost-Similar Relations. $b$ and $aBut$ are exactly-similar, $myBut$ is almost-similar to $b$.}
\label{typeusage-example}
    \scriptsize{
    \begin{minipage}{6cm}
    \begin{lstlisting}[language=java,boxpos=c, frame=single] 
    class A extends Page {
      Button createButton() {
        Button b = new Button();
        ...(interlaced code)
        b.setText("hello");
        ...(interlaced code)
        b.setColor(GREEN);
        return b;
      }
    }
    \end{lstlisting}
    \end{minipage} \hspace{.6cm} \begin{minipage}{6cm}
    \begin{lstlisting}[language=java,boxpos=c, frame=single]
    class B extends Page {
      Button void createButton() {
        ... (code before)
        Button aBut = new Button();
        ...
        aBut.setText("great");
        aBut.setColor(RED);
        return aBut;
      }
    }
    \end{lstlisting}
    \end{minipage} 
    
    \begin{minipage}{6cm}
    \begin{lstlisting}[language=java,boxpos=c, frame=single]
    class C extends Page {
      Button myBut;
      Button void createButton() {
        myBut = new Button();
        myBut.setColor(PINK)
        nyButton.setText("world");
        myBut.setLink("http://bit.ly");
        ... (code after)
        return myBut;
      }
    }
    \end{lstlisting}
    \end{minipage}
    }
\end{figure}

Those relations can be formalized as follows:

\begin{definition}
The relation \emph{exactly-similar} (noted $E$) is a relation between two type-usages $x$ and  $y$ of object-oriented software if and only if:
\begin{align*}
xEy \iff & T(x)=T(y)\\
&\land  C(x)=C(y)\\
&\land M(x)= M(y)
\end{align*}
\end{definition}

We also define for each type-usage $x$ the set $E(x)$ of all exactly-similar type-usages:
$E(x) = \{y | xEy\}$.

\begin{definition}
The relation \emph{almost-similar} (noted $A$) is a relation between two type-usages $x$ and  $y$ if and only if:
\begin{align*}
xAy \iff & T(x)=T(y)\\
&\land  C(x)=C(y)\\
&\land M(x)\subset M(y)\\
&\land |M(y)| = |M(x)|+ 1
\end{align*}
\end{definition}

For each type-usage $x$ of the codebase, the set $A(x)$ of almost-similar type-usages is noted:
\begin{align*}
A(x) = \{y | xAy\}
\end{align*}

It is possible to parameterize the definition of almost-similarity by allowing a bigger amount of difference, i.e. $|M(y)| = |M(x)|+k, k\geq 1$. However, our intuition is that k=1 is the best value because developers are more likely to write and commit code with a small deviation to standard usage. Indeed, committing code with large deviations from correct would produce more visible bugs much faster. We will also present empirical evidence supporting this assumption in \ref{influence-k}.

Furthermore, we can see that we obtain no \emph{exactly-similar} or \emph{almost-similar}
type-usages if the set of methods are not used in similar contexts.
Hence, for our approach to be applicable, for a given type, we need to have several usages of the type in the same context.
Consequently, we define a ``redundancy'' relation between type-usages. Two type-usages are said redundant if they have the same type and the same context. We will see in the evaluation that this redundancy condition is largely met in real software.

Finally, note that computing $E(x)$ and $A(x)$ for a type-usage $x$ with respect to a given codebase $Y$ of $n$ type-usages is done in linear time $O(n)$.

\subsection{S-score: A Measure of Strangeness for Type-usages}
\label{def:sscore}
Our approach is grounded on the assumption that the majority rule holds for type-usages too, i.e. that a type-usage is deviant if:
1) it has a small number of other type-usages that are \emph{exactly-similar}.
and 2) it has a large number of other type-usages that are \emph{almost-similar}.
Informally, a small number of \emph{exactly-similar} means ``only few other places do the same thing'' and a large number of \emph{almost-similar} means ``the majority does something slightly different''.  Assuming that the majority is right, the type-usage under inspection seems
deviant and may reveal an issue in software.

Now, let us express this idea as a measure of strangeness for type-usages, which we call \emph{S-score}. This measure will allow us to order all the type-usages of a codebase so as to identify the strange type-usages that are worth being manually analyzed by a software engineer.

\begin{definition}
  The S-score is:
  \begin{align*}
  \mbox{S-score}(x) = 1-\frac{|E(x)|}{|E(x)|+|A(x)|}
  \end{align*}
\end{definition}

This definition correctly handles extreme cases: if there are no exactly-similar type-usages and no almost-similar type-usages for a type-usage $a$, i.e. $|E(a)|=1$ ($E(x)$ always contains $x$ by definition) and $|A(a)|=0$, then $\mbox{S-score}(a)$ is zero, which means that a unique type-usage is not a strange type-usage at all. On the other extreme, consider a type-usage $b$ with $|E(b)|=1$ (no other similar type-usages) and $|A(b)|=99$ (99 almost-similar type-usages). Intuitively, a developer expects that this type-usage is very strange, may contain a bug, and should be investigated. The corresponding S-score is $0.99$ and supports the intuition.

\subsection{Predicting Missing Method Calls}
\label{predicting-the-missing}

Let us consider a strange type-usage $x$ (i.e. with at least one almost-similar type-usage).
Once A(x) is computed, we compute missing method call predictions as follows.
First, we collect the set $R$ of all calls that are present in almost-similar type-usages but missing in $x$. In other terms:

\begin{align*}
R(x) = \{ m | m \notin M(x) \land  m \in \bigcup _{z\in A(x)}{M(z)} \}
\end{align*}

For each recommended method in $R(x)$, we compute a likelihood value $\phi(m,x)$. The likelihood is the frequency of the missing method in the set of almost-similar type-usages: 

\begin{align*}
\phi(m,x) = \frac{|\{ z | z \in A(x) \land  m\in M(z) \}|}{|A(x)|}
\end{align*}

Eventually, predicting missing method calls for a type-usage $x$ consists of listing calls that are in $R$, and that have a likelihood value greater than a threshold $t$. 
\begin{align*}
missing(x,t) = \{ m | m \in R(x) \land  \phi(m,x)>t  \}
\end{align*}

For illustration, consider the example in figure \ref{likelihood-example}. The type-usage under study is $x$ of type \code{Button}, it has a unique call to the constructor. There are 5 almost-similar type-usages in the source code ($a$, $b$, $c$, $d$, $e$). They contain method calls to \code{setText} and \code{setFont}. \code{setText} is present in 4 almost-similar type-usages out of a total of 5. Hence, its likelihood is $4/5=80\%$. For \code{setFont}, the computed likelihood is 20\%.
Given a threshold of 75\%, the system recommends to the developer \code{setText} as a missing method call.
Finally, we emphasize that $t$ is the only tuning parameter of the overall approach. The sensitivity of this parameter will be studied in the next section.

Note that the prediction of missing method calls is based upon the majority rule (the majority being $A(x)$), hence it is only effective to detect missing calls in the types that are extensively used in the source code. If a method call is missing in a type-usage of a rarely used type or in an uncommon context, our approach is not effective.

\begin{figure}
\caption{An example computation of the likelihoods of missing method calls}
\label{likelihood-example}
\begin{minipage}{2.4in}
\begin{align*}
T(x)=&Button\\
M(x)=&\{<init>\}\\
A(x)=&\{ a,b,c,d \}\\
M(a)=&\{<init>, setText \}\\
M(b)=&\{<init>, setText \}\\
M(c)=&\{<init>, setText \}\\
M(d)=&\{<init>, setText \}\\
M(e)=&\{<init>, setFont \}\\
\end{align*}
\end{minipage}\hfill \begin{minipage}{2.4in}
$$R(x)={setText, setFont}$$
$$\phi(setText)=\frac{4}{5}=0.80$$
$$\phi(setFont)=\frac{1}{5}=0.20$$
\end{minipage}
\end{figure}

\section{Quantitative Evaluation}
\label{evaluation}
 
This section gives experimental results to validate our approach to detecting missing method calls.
It combines different quantitative techniques to validate the approach from different perspectives:
\begin{itemize}
  \item We show that our approach could help in Sandra's case (presented in \ref{helping-sandra}).
  \item We show that in real-software (presented in \ref{datasets}) (a) the S-score is low for most type-usages of real software, i.e. the majority of real type-usages is not strange (see \ref{descriptive-stats}), and (b) the S-score is able to catch type-usages with a missing method call, i.e. that the S-score of such type-usages is on average higher than the S-score of normal type-usages (see \ref{automatic-evaluation}).
  \item We show that our algorithm is able to predict missing method calls that are actually missing (see \ref{automatic-guessing}).
\end{itemize}

\subsection{Helping Sandra}
\label{helping-sandra}

Let us go back to Sandra's problem presented in \ref{sandra}. 
If one assumes that Sandra makes no call at all on \emph{this}, then the almost similar type-usages are those of type \code{DialogPage}, in the context of \code{createControl}, with one single method call on \emph{this}.
In the Eclipse codebase v3.5, there are 16 such type-usages, all containing a single call to \code{setControl} (the other instances of \code{DialogPage} have more calls on this).
Hence, in this case, the system predicts a S-score of $16/17=0.94$, which means very strange, and recommend a call to \code{setControl} with a likelihood of 100\%.
Eventually, based on the high S-score value and the high likelihood value, Sandra adds the missing method call, solves her bug, and continues her implementation task.

However, note that our technique provides the name of the missing method but not its parameters or its relative position in code. Hence, before adding the call, Sandra has to read the appropriate documentation about the parameters of \code{setControl}, and to decide where to place the missing method call. Predicting the exact location (e.g. the line number of the missing method call) or the parameters to the missing call are beyond the scope of this paper.

\subsection{Datasets}
\label{datasets}

We have implemented an extractor of type-usages for Java bytecode using the Soot bytecode analysis toolkit \cite{Vallee-Rai1999}. In Appendix \ref{replication}, replication guidelines are given (including datasets and software). In all, we mined 5 different large-scale datasets:
\begin{description}
\item[eclipse-swt] All type-usages of the Eclipse development environment version 3.5 related to SWT types (SWT is the graphical user-interface library underlying Eclipse)
\item[eclipse-java.util] All type-usages of the Eclipse development environment version 3.5 related to standard Java types from package \code{java.util}.
\item[eclipse-java.io]  All type-usages of the Eclipse development environment version 3.5 related to standard Java types from package \code{java.io}.
\item[apache-tomcat] All type-usages of the Apache Tomcat web application server related to domain types (i.e. \code{org.apache.tomcat.*}).
\item[apache-derby] All type-usages of the Apache Derby database system related to domain types (i.e. \code{org.apache.derby.*}).
\end{description}
The programs pairs were selected to cover different application domains (IDE, web server, database) and libraries (GUI, IO).

\begin{table}
\caption{Descriptive Statistics of the Datasets. They all have an order of magnitude of $10^4$ type-usages.}
\label{table:datasets}

  \centering\begin{tabularx}{\textwidth}{|p{2cm}| Y  | Y | Y | Y |} 
  \hline
  Datasets          & \#types & \#contexts & \#type-usages & \#redundant            \\
  \hline
  eclipse-swt       & 389     & 7839       & 41193         &   28306 (68\%)         \\
  eclipse-java.io   & 54      & 3300       & 9765          &   6820  (70\%)         \\
  eclipse-java.util & 70      & 14536      & 40251         &   21780 (54\%)         \\
  derby             & 1848    & 10218      & 36732         &   11164 (30\%)         \\
  tomcat            & 1355    & 3891       & 18904         &   8467  (45\%)         \\
  \hline
  \end{tabularx}
\end{table}

Table \ref{table:datasets} presents descriptive statistics for the datasets. The first column gives the number of different types in the datasets and the second the number of different contexts (the number of different method signatures, see \ref{type-usages}).
For instance, in the Eclipse codebase, there are 389 different types from SWT that are used in 7839 different methods.
The third column gives the overall number of type-usages extracted and the last one the number of redundant type-usages (at least one other type-usage with same context and same type, see \ref{relations-for-type-usages}) associated with the ratio of redundant type-usages. 
Those statistics support the following interpretations:

First, the shape of each dataset is quite different, both in terms of types and in terms of contexts.
For instance, the number of types of eclipse-java.io is only 54 because this library is used for input input/output, whose logic can be encapsulated in a few types (e.g. \code{File} and \code{PrintStream}).
On the contrary, the whole application logic of the Derby database system is spread over 1848 different types.
Hence, the more complex the logic, the more classes are used to express it.

Second, we note that the distributions of type-usages per type and type-usages per context follow a power-law distribution, which means that a few types and a few contexts trust a large number of type-usages.
For instance, for eclipse-swt, the top-20 most popular types (out of 389) cover 62\% of the type-usages.
This goes along the same line as the results of Baxter et al. \cite{Baxter2006}.

Third, since our approach requires equality of type-usage contexts (the signature of the enclosing method) to build the set of almost-similar type-usages, it is applicable only when one can observe at least two type-usages for a given context (this is the key part of the definition of redundant type-usages).
Since the number of redundant type-usages is high (up to 68\% as shown by the last column of table \ref{table:datasets}), the overall applicability of our approach is validated.

\subsection{The Correctness of the Distribution of the S-score}
\label{descriptive-stats}

For each dataset and for each type-usage, we have computed the sets of exactly and almost-similar type-usages ($E(x)$ and $A(x)$) and the corresponding S-score.
Since all datasets are extracted from widely used and mature software, we assume that most of the type-usages have a low degree of strangeness, i.e. a low S-score.

\begin{figure}[ht]   
  \caption{Distribution of the S-score based on the type-usages of type SWT.* in the Eclipse codebase. Most type-usages have a low S-score, i.e. are not strange.}
  \label{fig:distribution-sscore}
  \centering\includegraphics[width=\textwidth]{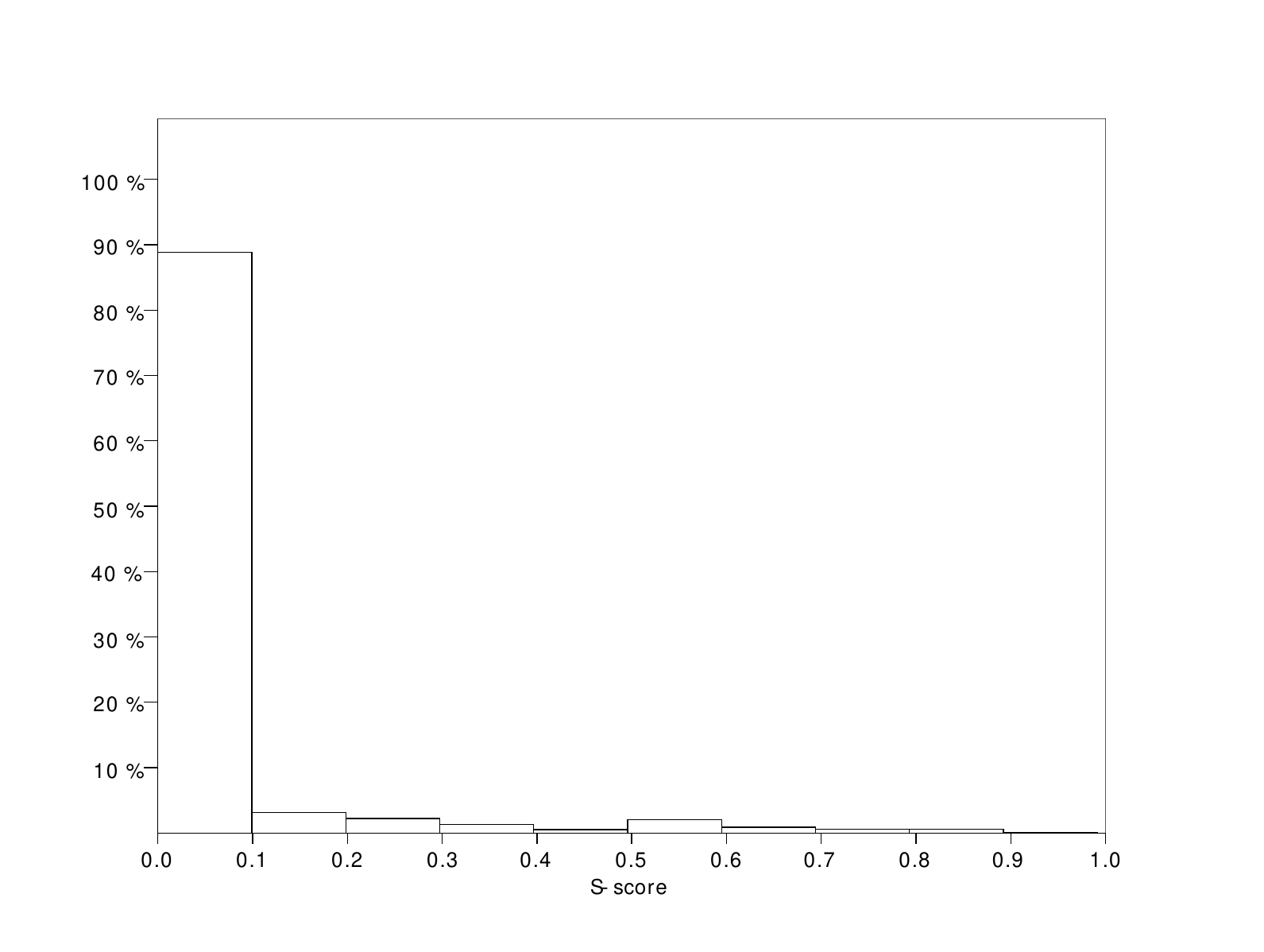}
\end{figure}

Table \ref{tab:distributions} validates the assumption: the median S-score is 0 for all datasets (i.e. more than 50\% of all type-usages have a S-Score of 0).
Furthermore, there is always a very small proportion of strange and very strange type-usages (S-score>0.5 and 
S-score>0.9). 
Figure \ref{fig:distribution-sscore} shows the complete distribution of the S-score for dataset \emph{eclipse-swt}. This figure graphically shows that a large majority of type-usages has a low S-score and that the distribution is exponential. This kind of distribution shape holds for all datasets.

\begin{table}
\caption{Distribution of the S-score on different datasets}
\label{tab:distributions}
  \centering\begin{tabularx}{\textwidth}{|p{2cm}| >{\centering}p{.8cm} | >{\centering}p{.8cm} | >{\centering}p{.8cm} | Y | Y | Y |} 
  \hline
  Datasets         &\scriptsize{\#type-usages}& \scriptsize{Median S-score}  & \scriptsize{Mean S-score} & \scriptsize{S-score<0.1} & \scriptsize{S-score>0.5} & \scriptsize{S-score>0.9} \\
  \hline
  eclipse-swt      &   41193     & 0              & 0.04         &   89\%      & 2.7\%       & 0.1\%       \\
  eclipse-java.io  &   9765      & 0              &    0.03      &   92\%      &   0.7\%     &       0\%   \\
  eclipse-java.util&   40251     & 0              &    0.02      &   94\%      &   0.8\%     &      0.003\%\\
  derby            &   36732     & 0              & 0.01         &   98\%      &   0.3\%     & 0\%         \\
  tomcat           &   18904     & 0              & 0.01         &   98\%      &   0.1\%     & 0\%         \\ 
  \hline
  \end{tabularx}
\end{table}

\subsection{The Ability of S-score to Catch Degraded Code}
\label{automatic-evaluation}

\begin{lstlisting}[language=perl, numbers=left, boxpos=c, frame=single, numberstyle=\tiny, float,caption={An Algorithm to Simulate Missing Method Calls},label=alg:degraded]
$L = {} // the set of degraded type-usages
foreach type-usage $t
  check that $t has redundant type-usages
  foreach method-call $m of $t
    $o = clone($t)
    remove $m from $o
    add $o to $L
  end
end
return $L
\end{lstlisting}

Now, we show that a faulty type-usage would be caught by the S-score (faulty in the sense of suffering from a missing method call). For this to be true, a type-usage with a missing method call should have a higher S-score than a normal type-usage.

To assess the validity of this assumption, our key insight is to simulate missing method calls.
Given a type-usage from real software with $n$ method calls, our idea is to create $n$ degraded type-usages by removing one by one each method call.
This strategy to create validation input data has several advantages:
(a) there is no need to manually assess whether a type-usage is faulty, we know it by construction, (b) it is based on real data (the type-usages come from real software) and (c) it yields a large-scale evaluation with a large number of evaluation cases.

Listing \ref{alg:degraded} presents this algorithm as pseudo-code. Two for-loops traverse the whole dataset for each type-usage and for each method calls. Line 3 filters the type-usages that have no redundant type-usages. The rationale of this filtering is that if a type-usage is already the only one of a given type in a given context, i.e. is already an outlier in the dataset, it does not make sense to degrade it further (because by construction, all resulting degraded type-usages would have no almost-similar type-usages).

Once we have a set of artificial degraded type-usages, we can analyze the distribution of their S-score and compare it with the distribution of real type-usages computed in \ref{descriptive-stats}.
Although the S-score has been designed to reveal missing method calls, it may be inefficient due to the peculiarities of real data. The comparison enables us to assess that the S-score is actually well-designed.

We have conducted this evaluation for all datasets. Table \ref{distrib-normal-vs-degraded} gives the results.
For all datasets, the median S-score of degraded type-usages is 1 (recall that a S-score of 1 means that there is no exactly-similar type-usages) and the mean S-score always greater than 0.60. A significant proportion of degraded type-usages has a S-score greater than 0.9 (from 60\% for \emph{derby} to 78\% for \emph{tomcat}).
Also, there is always some degraded data that cannot be recognized as problematic (S-score lesser than 0.1), up to 38\% for \emph{derby}.

Finally, let us discuss the number of simulated missing method calls (the first column in table \ref{distrib-normal-vs-degraded}). Since there is on average two method calls per type-usage, the number of simulated missing calls should be of the same order of magnitude as two-times the number of redundant type-usages. The comparison of table \ref{table:datasets} and table \ref{distrib-normal-vs-degraded} validates this assumption. For instance, there are 22673 redundant type-usages in \emph{eclipse-swt} and 42845 degraded type-usages resulting from algorithm \ref{alg:degraded}.

\begin{table}
\caption{Distribution of the S-score of Degraded Type-usages resulting from Algorithm \ref{alg:degraded}. The S-score is able to capture faulty type-usages.}
\label{distrib-normal-vs-degraded}
  \centering\begin{tabularx}{\textwidth}{|p{2cm}| p{2cm} | Y | Y | Y | Y | Y |} 
  \hline
  Datasets         &\#simulated missing calls& Median S-score & Mean S-score & S-score<0.1 & S-score>0.5 & S-score>0.9 \\
  \hline
  eclipse-swt      &     42845      &    1           &      0.78    &   18\%      &  79\%       &  73\%       \\
  eclipse-java.io  &     9698       &    1           &      0.69    &   30\%      &    68\%     &      67\%   \\
  eclipse-java.util&     34049      &    1           &      0.76    &   22\%      &    76\%     &         74\%\\
  derby            &     16254      &    1           &      0.61    &   38\%      &   61 \%     &60\%         \\
  tomcat           &     10589      &    1           &      0.78    &   22\%      &   78 \%     &78\%         \\ 
  \hline
  \end{tabularx}
\end{table}

\subsection{The Performance of Missing Method Calls Prediction}
\label{automatic-guessing}
The third evaluation of our system measures its ability to guess missing method calls.
The assumption underlying this evaluation is that our approach to detect missing method calls (presented in \ref{predicting-the-missing}) should be able to predict calls that were artificially removed.

For this, we have used the same setup used for evaluating the characteristics of the S-score (see \ref{automatic-evaluation}) but instead of looking at the distribution of the S-score of degraded data, we have tried to guess the method call that was artificially removed.

For instance, given a real type-usage of the codebase representing a \code{Button} and containing \code{<init>} and \code{setText}, we test the system with two different queries: 1) \code{<init>} only and 2) \code{setText} only.  The system may predict several missing method calls, but a perfect prediction would be \code{setText} as missing method call for the first query and \code{<init>} for the second query.

Hence, for each dataset, the system is evaluated with the same number of queries as in \ref{automatic-evaluation}. Then, we collect the following indicators for each query:
$correct_i$ is true if the removed method call is contained in the set of predicted missing method calls;
$answered_i$ is true if the system predicts at least one missing method call (we recall that the system predicts a missing method call if there is at least one almost-similar type-usage);
$sizeanswer_i$ is the number of predicted missing method calls;
$prefect_i$ is true if the system predicts only one missing method call and it's the one that was removed, i.e. $perfect_i \implies correct_i \land (answered_i=1)$.

For N queries, we measure the overall performance of the system using the following metrics:
\begin{itemize}
\item {\bf ANSWERED} is the percentage of answered queries. A query is considered as answered if the system outputs at least one missing method call. $$ANSWERED=\frac{|\{i|answered_i\}|}{N}$$
\item {\bf CORRECT} is the percentage of correctly answered queries.
\item {\bf FALSE} is the percentage of incorrectly answered queries, i.e. queries without the removed method call in the predicted missing ones.
\item {\bf PRECISION} is the classical information retrieval precision measure. Since the precision is not computable for empty recommendations (i.e. unanswered queries), $$PRECISION=\frac{\sum _{i|correct_i} 1/sizeanswer_i}{|\{i|answered_i\}|}$$
\item {\bf RECALL} is the classical information retrieval recall measure: $$RECALL=\frac{|\{i|correct_i\}|}{N}$$

\end{itemize}

For each dataset, we evaluated the \toolname~system based on the evaluation process and performance metrics presented above. Table \ref{result-eval-predicting-missing} presents the results.
The system is able to answer from 54\% to 76 \% of the queries depending on the dataset, and when it answers, it's usually correct (CORRECT varies from 73\% to 89\%).
Furthermore, the correct missing method call is not lost among a large number of wrong predictions, since the PRECISION value remains high, from 59\% to 83\%.

The precision and recall values presented here are lower than those presented in our conference paper \cite{Monperrus2010a} because we have changed the evaluation strategy. In the previous version of the evaluation algorithm which simulates missing method calls, we discarded type-usages that have only one call to be the seed of degraded type-usages (meaning we discarded all artificial queries with no calls). We discovered that those queries are actually the most difficult to answer, yielding to lower precision and recall. Since it's often the case that real type-usages contain no call (objects simply passed as parameter to other methods),
our new evaluation strategy is more realistic, but also decreases the measured performance.

\begin{table}
\caption{Performance metrics the \toolname~system for different datasets for threshold $t=0.9$.}
\label{result-eval-predicting-missing}
\centering\begin{tabularx}{\textwidth}{|p{2cm}|X|X|X|X|X|X|}
\hline
Dataset          & $N_{query}$ & ANSWERED &  CORRECT & FALSE  & PRECISION     &     RECALL \\
\hline
eclipse-swt      &   42845     &   76\%   &   89\% &  11\%   &     77\%      &   68\%     \\
eclipse-java.io  &   9698      &   66\%   &   87\% &  13\%  &     83\%      &   58\%     \\
eclipse-java.util&   34049     &   75\%   &   98\% &  12\%  &     81\%      &   66\%     \\
derby            &   16254     &   54\%   &   76\% &  24\%  &     72\%      &   41 \%     \\
tomcat           &  10589      &   67\%   &   73\% &  27\%  &     59\%      &   49\%     \\
\hline
\end{tabularx} 
\end{table}

\subsection{Influence of the Filtering Threshold}
\label{influence-filtering}
As said in \ref{predicting-the-missing}, our approach is only dependent on one single tuning parameter: the threshold value to recommend missing method calls.
In this section we study the sensitivity of the performance of our system with respect to this parameter.

For the dataset \emph{eclipse-swt}, we have computed all performance metrics for different values of the threshold.
Table \ref{result-eval-threshold} gives the result of this experiment. We can see that all metrics vary in an interval of 2\% for CORRECT to 10\% for PRECISION.
We consider that this small variance shows that the our system is not crucially sensitive to this tuning parameter.
Note that the threshold -- by removing some method calls from the recommendations -- mechanically decreases ANSWERED, CORRECT and RECALL. Logically, the higher the threshold is, the higher the precision.
Interestingly, setting the threshold to extreme values (0 or 1) is neither very good for the precision nor very bad for the recall; this shows that much of the performance of the approach comes from the definition of the exact-similarity and almost similarity themselves.

In this paper, we use the threshold for sake of evaluation,
but in practice, developer actually inspect the recommended missing method calls in descending order of likelihood to be missing.

\begin{table}
\caption{Experimental results on the influence of the filtering threshold for dataset \emph{eclipse-swt}.}
\label{result-eval-threshold}
\centering\begin{tabularx}{\textwidth}{|p{2cm}|X|X|X|X|X|X|}
\hline
Threshold          & $N_{query}$ & ANSWERED &  CORRECT & FALSE  & PRECISION     &     RECALL \\
\hline
0      &   42845     &   82\%   &   91\% &  9\%   &     67\%      &   74\%     \\
0.2    &   42845      &   82\%   &   90\% &  10\%  &     68\%      &   73\%     \\
0.4    &   42845     &   81\%   &   89\% &  11\%  &     71\%      &    72\%     \\
0.6    &  42845      &   78\%   &   88\% &  12\%  &     74\%      &   69\%     \\
0.8    &  42845      &   77\%   &   88\% &  12\%  &     76\%      &   68\%     \\
0.9    &   42845     &   76\%   &   89\% &  11\%   &     77\%      &   68\%     \\
1      &   42845     &   76\%   &   89\% &  11\%  &     77\%      &   68\%     \\
\hline
\end{tabularx} 
\end{table}

\subsection{Influence of the Context Equality Condition}
\label{influence-context}

While we saw in \ref{influence-filtering} that the filtering threshold is not crucial for the system performance, we now evaluate the importance of our context definition with respect to precision and recall. For all datasets, we apply exactly the same evaluation strategy as in \ref{automatic-guessing}, keeping all settings equal but removing the necessary condition of context equality that we defined for exact-similarity and almost similarity (see \ref{relations-for-type-usages}).

Table \ref{tab:influence-context} shows the results of this evaluation. 
Adding the context equality condition in our approach has a significant impact on both precision and recall.
For instance, let us consider the eclipse-java.io dataset, the precision of the system increases from 33\% to 83\% if we add the context equality condition in our algorithm.
An explanation is that Java.io classes (e.g. \code{FileInputStream}) highly depend on the context (for example, one usually does not open and close a file stream in the same method); hence taking the context into account highly increases the precision.

Overall, this experiment shows that the context equality condition is a crucial part of our algorithm: removing the context introduces a lot of noise (not relevant items in $A(x)$ and $E(x)$ in the predicted missing method calls) and hampers both the precision and the recall of the system.
This result is an empirical proof that the signature of the enclosing method largely conditions correct API usages. This key ingredient of your approach enables us to achieve a low level of false positives. 

\begin{table}
\caption{Experimental results on the influence of the context definition.}
\label{tab:influence-context}
\centering\begin{tabularx}{\textwidth}{|p{2cm}|X|X|X|X|}
\hline
Dataset          & PRECISION (w/o context) & PRECISION (with context)  & RECALL (w/o context)   &     RECALL (with context) \\
\hline
eclipse-swt      &      53\%    &  77\%  &     69\%      &   68\%     \\
eclipse-java.io  &      33\%    &  83\%  &     75\%      &   58\%     \\
eclipse-java.util&      42\%    &  81\%  &     90\%      &   66\%     \\
derby            &      43\%    &  72\%  &     40\%      &   41 \%     \\
tomcat           &      35\%    &  59\%  &     62\%      &   49\%     \\
\hline
\end{tabularx} 
\end{table}

\subsection{Influence of the Number of Method Calls in the Definition of Almost-Similarity}
\label{influence-k}

In the definition of almost-similarity presented in \ref{relations-for-type-usages}, we state that a type-usage is almost-similar to another one if and only if it contains a single additional method call. In this section, we evaluate the effect of considering more additional method calls to build the set of almost-similar type-usages. In the following, $k$ refers to the number of additional method calls.

For the dataset \emph{eclipse-swt}, and the evaluation strategy of \ref{automatic-guessing}, we have computed the following metrics for different values of $k$:
\begin{itemize}
\item The average number of exactly-similar type-usages, i.e. the average size of $E(x)$. This is independent of k and should remain constant. 
\item The average number of almost-similar type-usages, i.e. the average size of $E(A)$. The larger k, the more elements in $A(x)$, so this should increase with k.
\item The average S-score. By construction, since $|E(x)|$ is constant and $|A(x)|$ increases, it should increase too.
\item The average number of methods present in elements of $A(x)$ but absent in $E(x)$, i.e. the number of potentially missing method calls (noted $|R(x)|$ in \ref{predicting-the-missing}). Since $A(x)$ is bigger when k increases, it should increase too.
\item The average $\phi(m,x)$ of missing method calls. By construction of the formula of $\phi$, it should decrease when k increases.
\item The average number of recommended missing method calls after filtering on $\phi(m,x)$. We could not predict the impact on this number, because it results from two opposite effects on the number of potentially missing method calls (that increases) and on $\phi(m,x)$ (that decreases).
\item The precision and recall as described previously.
\end{itemize}

Table \ref{result-eval-influence-k} gives the result of this experiment.
First, all our assumptions are validated: the average $|E(x)|$ is constant, the average $|A(x)|$, S-score, and number of potentially missing method calls $|R(x)|$ increases with k, and the average $\phi(m,x)$ decreases with k.
Interestingly, the average number of recommended missing method calls after filtering on $\phi(m,x)$ increases with k. It means that the effect of k on $|R(x)|$ is more important than the effect on $\phi$, and results in an increase of recommended missing method calls.
We also see that the precision decreases with k. This is due to the larger number of recommended method calls. As we assumed in \ref{relations-for-type-usages}, and according to this evaluation setup, the optimal value for k is 1.
A remarkable effect is that the recall does not increase with k. This is due to the filtering effect described above.
The new missing method calls discovered in the neighborhood when increasing k have a low $\phi(m,x)$, and they remain under the threshold to be predicted as missing.

\begin{table}
\caption{Experimental results on the influence of the number of additional method calls $k$ (in the definition of almost-similarity). Dataset considered: \emph{eclipse-swt}, $missing(x,t) = \{m|\phi(m,x)>0.9\}$.
k=1 achieves the best precision, with no loss in recall.}
\label{result-eval-influence-k}
\centering\begin{tabularx}{\textwidth}
{|X  |X          |X         | X         |X        |X               |X                |X          |X        |}
\hline
k    &Avg |E(x)| &Avg |A(x)|&Avg \mbox{S-score}&Avg |R(x)|&Avg $\phi(m,x)$&Avg |Miss(x)| & Precision & Recall \\
\hline
1    & 1.75      & 38.75    & 0.78      & 1.85     & 0.71          & 1.03            &  77\%     &  68\%  \\
2    & 1.75      & 45.18    &0.85       & 2.86     & 0.67          & 1.22            &  70\%     &  69\%  \\
3    & 1.75      & 48.22    & 0.87      &3.46      &0.67           &1.27             &  69\%     &  68\%  \\
4    & 1.75      & 49.11    & 0.87      &3.75      &0.33           &1.28             &  68\%     &  68\%  \\
\hline
\end{tabularx} 
\end{table}

\section{Qualitative Evaluation}
\label{case-studies}

The evaluation results presented in \ref{automatic-evaluation} and \ref{automatic-guessing} suggest that a software engineer should seriously consider analyzing a type-usage if it has a high S-score.
However, it may be the case that our process of artificially creating missing method calls does not reflect real missing method calls that occur in real software.

As a counter-measure to this threat of validity, we used the \toolname~system in a real-world setting. 
We searched for missing method calls in the Eclipse, Apache Tomcat and Apache Derby software packages.
Finding missing method calls in those packages is ambitious for the following reasons:
\begin{itemize}
\item since the community of users is large, the software is used daily in plenty of different manners, and missing method calls have a chance to produce a strange behavior and hence, are likely to be already repaired.
\item since the community of developers is large and the codebase is several years old, most of the code has been read by several developers, which increases the probability of detecting suspicious code.
\end{itemize}

We analyzed approximately 30 very strange type-usages - 30 corresponding to approximately 3 full days of analysis, because we were totally unfamiliar with the Eclipse code base.
For each of them, we analyzed the source code in order to understand what the high S-score means. Then, we tried to formulate the problem as an issue report and to write a patch that solves the issue.

For example, let us consider issue \#326504\footnote{\scriptsize\url{https://bugs.eclipse.org/bugs/show_bug.cgi?id=326504}} in Eclipse.
The system reports a strange type-usage (S-score: 0.93, \#almost-similar: 109) in NewNodeDialog.create\-Dialog\-Area related to a composite. By reading the source code, it turns out that this method does not comply with an API best practice. Then we reported the issue in natural language (\emph{``There is a muddle of new, super and returned Composite in
NewNodeDialog.createDialogArea''}) and we attached a patch presented in listing \ref{patch:NewNodeDialog}. 

\begin{lstlisting}[numbers=left, boxpos=c, frame=single, float, basicstyle=\scriptsize, numberstyle=\tiny, caption={Patch submitted to the Eclipse Bug Repository to Solve a Strange Type-Usage (issue number \#326504)},label=patch:NewNodeDialog] 
--- src/org/eclipse/equinox/.../NewNodeDialog.java  18 Apr 2008
+++ src/org/eclipse/equinox/.../NewNodeDialog.java  29 Sep 2010
@@ -47,8 +47,7 @@
  }
 
  protected Control createDialogArea(Composite parent) {
-   Composite compositeTop = (Composite) super.createDialogArea(parent);
-   Composite composite = new Composite(compositeTop, SWT.NONE);
+   Composite composite = (Composite) super.createDialogArea(parent);
 
    setMessage(SecUIMessages.newNodeMsg);
\end{lstlisting}

In our previous report on this subject \cite{Monperrus2010a}, we tried to present quantitative arguments (such as the ratio of false positives) from this study.
With hindsight, we find today that those quantitative arguments were artificial because the term ``false positive'' is too strict in this evaluation context. As discussed in much detail below, a strange type-usage may have different meanings: 
\begin{itemize}
\item the missing method call has to be added (e.g.  discussion in \ref{VAPIBP});
\item the containing method smells, it could be rewritten or refactored (e.g. discussion in \ref{encapsulation-breaking});
\item the missing method call highlights a weakness of the API under consideration (e.g. discussion in \ref{FP}) 
\end{itemize}
We now think that these kinds of strange type-usages cannot be meaningfully summarized with a false/true positive classification. Hence, we rather discuss our findings about the issues that a tool like ours reveals in software.

\begin{table}
\caption{Issues reported based on the predictions of our system.}
\label{table:issues}
\centering\begin{tabularx}{\textwidth}{|X|X|}
\hline
Issue identifier & Outcome \\
\hline
Eclipse issue \#296552 &  validated, patch accepted\\ 
Eclipse issue \#297840 &  no answer\\ 
Eclipse issue \#296554 &  code no longer maintained\\ 
Eclipse issue \#296586 &  wrong analysis\\ 
Eclipse issue \#296581 &  validated, patch accepted\\ 
Eclipse issue \#296781 &  validated, patch accepted\\ 
Eclipse issue \#296782 &  validated, patch accepted\\ 
Eclipse issue \#296578 &  no answer\\ 
Eclipse issue \#296784 &  validated, patch accepted\\ 
Eclipse issue \#296481 &  validated, patch accepted\\ 
Eclipse issue \#296483 &  validated, patch accepted\\ 
Eclipse issue \#296568 &  validated\\ 
Eclipse issue \#275891 &  validated\\ 
Eclipse issue \#296560 &  code no longer maintained\\ 
Eclipse issue \#326504 &  validated, patch accepted\\ 
Tomcat issue \#50023   &  wrong analysis\\
Derby issue \#DERBY-4822 &  validated, patch accepted\\ 
\hline
\end{tabularx} 
\end{table}

\subsection{Finding: The S-score enables developers to find issues in mature software}

Table \ref{table:issues} presents our issue reports that are based on the predictions of our system.
In all, we reported 17 issues and got positive feedback for 11 of our reports, including 9 accepted patches that are now in the HEAD version of the corresponding revision control system.
This shows that the S-score is able to reveal issues in mature software.
To really appreciate these numbers, consider this question: what is the probability of submitting a valid patch on a very large unknown codebase (between $10^5$ and $10^6$ lines of code)?
We believe that the answer is ``very low''.
However, with tool support, we were able to do so.
Our system pointed us directly to very strange pieces of code, and gave us sufficient information (what method calls the majority does) to understand the issue and submit a patch.

\subsection{Finding: Missing method calls may reveal software aging}

Missing method calls may reveal problems related to software aging.
Let us explain this finding with two concrete cases.

According to the system, Eclipse's \code{ExpressionInputDialog} contains strange code related to closing and disposing the widgets of the dialog.
We reported this issue as issue \#296552
\footnote{see \scriptsize\url{https://bugs.eclipse.org/bugs/show_bug.cgi?id=296552}} and our patch was quickly accepted.
Interestingly, by mining the history of this class, we found a set of commits and a discussion around another issue report \footnote{see \scriptsize\url{https://bugs.eclipse.org/bugs/show_bug.cgi?id=80068}}.
Even if this other issue report was closed as \emph{solved}, the code was never cleaned and the measures and counter-measures taken during the discussion degraded the code quality and increased its strangeness.

\begin{lstlisting}[numbers=left, boxpos=c, frame=single, basicstyle=\scriptsize, numberstyle=\tiny,float,caption={Software Aging: Unnecessary Code Related to a Past Version of the API},label=ChangeEncodingAction]
Control composite= super.createDialogArea(parent);
// this check is not relevant anymore
if (!(composite instanceof Composite)) {    
        composite.dispose();    
        composite= new Composite(parent, SWT.NONE);
}
\end{lstlisting}

The other example is in \code{ChangeEncodingAction} which contains defensive code related to a very old version of the API (see listing \ref{ChangeEncodingAction}). This check is completely unnecessary with the current version. 
Our approach finds that having calls to \code{super.\-cre\-ateDialogArea}, \code{dispose} and \code{new Composite} is really strange for a type-usage of type \code{Composite}. Following our remark about this class to the Eclipse developers, the code has been actualized. In this case, software aging comes from changes of the API that were not reflected in client code.

\subsection{Finding: Missing method calls may reveal violations of API best practices}
\label{VAPIBP}
Strange type-usages often reveal violations of API best practices. An API best practice is a programming rule which is not enforced by the framework code or the programming language.
In the following, we discuss several violations of different API best practices of Eclipse which were detected by our system.

\textbf{Best practice - setting fonts:}

A best practice of Eclipse consists of setting the font of new widgets based on the font of the parent widget and not on the system-wide font. Not following this best practice may produce an inconsistent UI.
To our knowledge, this API best practice is not explicitly documented but pops up in diverse  locations such as:
newsgroups\footnote{\scriptsize\url{http://dev.eclipse.org/viewcvs/index.cgi/org.eclipse.ui.browser/src/org/eclipse/ui/internal/browser/BrowserDescriptorDialog.java}},
issue reports\footnote{\scriptsize\url{https://bugs.eclipse.org/bugs/show_bug.cgi?id=175069} and \url{https://bugs.eclipse.org/bugs/show_bug.cgi?id=268816}},
and commit texts\footnote{\scriptsize\url{http://dev.eclipse.org/viewcvs/index.cgi/org.eclipse.ui.ide/src/org/eclipse/ui/internal/ide/dialogs/ResourceInfoPage.java?sortby=log&view=log} and \url{http://dev.eclipse.org/viewcvs/index.cgi/org.eclipse.ui.browser/src/org/eclipse/ui/internal/browser/BrowserDescriptorDialog.java}}.

\begin{figure}[t]
  \caption{Excerpt of revision 1.5 of Apr. 10 2006 of \code{BrowserDescriptorDialog.java}. Two missing method calls related to setting fonts are added.}
  \label{fig:solve-font}
  \centering\includegraphics[width=\textwidth]{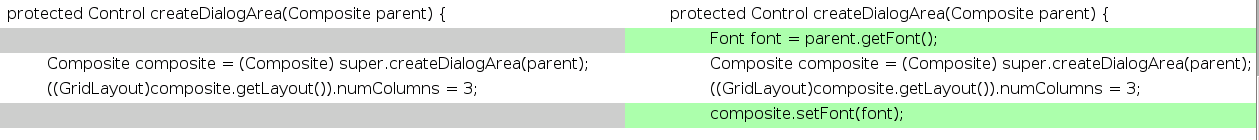}
\end{figure}

The programming rule associated to this API best practice is to call \code{getFont} on the parent widget and to call \code{setFont} on the newly created widget. Figure \ref{fig:solve-font} illustrates this point by showing the result of a commit which solves a violation of this best practice: the new code at the right hand side contains the previously missing method calls.
Our system automatically detects the missing calls related to such violations.

\textbf{Best practice - calling dispose:}

The SWT toolkit uses operating system resources to deliver native graphics and widget functionalities. While the Java garbage collector handles the memory management of Java objects, it cannot handle the memory management of operating system resources. Not disposing graphical objects is a memory leak, which can be harmful for long-running applications.
For instance, the following code of \code{ExpandableLayout} produces a high S-score (0.96):
\begin{verbatim}
// Location: ExpandableLayout.layout
size= FormUtil.computeWrapSize(new GC(..),..)
\end{verbatim}
The newly created graphical object (\code{new GC()}) is not assigned to a variable. However, the Java compiler inserts one in the Java bytecode. Since the method \code{computeWrapSize}, which receives the new object as a parameter, does not dispose the new object, it is never disposed. That's why our system predicts a missing call to \code{dispose}.
This problem was filed and solved in the Eclipse Issue repository independently of our work\footnote{\scriptsize\url{https://bugs.eclipse.org/bugs/show_bug.cgi?id=257327}}. 

\subsection{Finding: Missing method calls are sometimes due to software cloning}
\label{IC}
A common problem revealed by the system is incorrect cloning of code (already empirically identified by Thummalapenta et al. \cite{Thummalapenta2010}).
In a particular context, most of the type-usages of a given type uses a certain procedure to accomplish a task. Strange code uses a incorrect clone of the procedure code, i.e. a clone with some missing method calls.

For instance, it is standard to create the new container widget of a dialog using the framework method \code{createDialogArea} of the super class \code{Dialog}. This best practice is documented in the API documentation of \code{Dialog} (see Listing \ref{super.createDialogArea}).
The consequences of violating this API best practice are diverse: mainly UI weirdness and code that is hard to understand and maintain.

\begin{lstlisting}[numbers=left, boxpos=c, frame=single, basicstyle=\scriptsize, numberstyle=\tiny,float,caption={API Documentation of an API Best Practice Related to a Method Call},label=super.createDialogArea]
Dialog.java, line 780
   * Subclasses must override this method but may call super as
   * in the following example:
   * 
   * Composite composite = (Composite) super.createDialogArea(parent);
   * //add controls to composite as necessary
   * return composite;
\end{lstlisting} 

However, certain type-usages do not follow this API best practice and clones the functionality of the super implementation (of \code{Dialog.createDialogArea}).
However, there is sometimes an important method call present in the super implementation but missing in the duplicated functionality.

For instance, in \code{Add\-Source\-Con\-tainer\-Dialog} the new \code{Composite} is instantiated and incompletely initialized by hand, yielding to high S-score (that we reported in Eclipse issue \#296481, fixed).
Similarly, a composite \code{UpdateAndInstallDialog} is initialized using an internal private method and deserves a high S-score as well (Eclipse issue \#296554, fixed).
In both cases, the duplicated clone is not 100\% compliant with \code{Dialog.createDialogArea} that should be called.
A call that is present in the super implementation is missing in the clone and triggers a very high S-score.

To sum up, software cloning leads to slightly incorrect correct that is revealed by missing method calls.

\subsection{Finding: The core algorithm is sensible to the tyranny of the majority}
\label{FP}

The expression \emph{tyranny of the majority} is sometimes used to refer to political and decision systems in which minorities are not allowed to express their differences.
The concept applies very well to our system: if a type-usage has a small number of exactly-similar other type-usages and a high number of almost-similar, the type-usage is part of a minority (the exactly-similar type-usages) with respect to a large majority (the almost-similar type-usages).
By construction, the formula of the S-score always yields high values for minorities, even if they are actually correct.

For instance, there is a couple of cases in Eclipse where an empty \code{Label} and an empty \code{Composite} are used to create filler and spacers. Usually and conceptually, labels are used to contain a text (and host a call to \code{setText}) and composite are used to contain widgets organized with a layout strategy (and host a call to \code{setLayout}). As a result, an empty \code{Label} and an empty \code{Composite} always trigger high S-scores to the corresponding type-usages. To a certain extent, using an empty label or composite is more a hack than a good practice. A better solution would be, for instance, to configure the margins of the layout, or to introduce a \code{Filler} class in the SWT library.

In those cases, there is a conjunction of three factors:
\begin{itemize}
\item the lack in the considered API of a class dedicated to a particular concern (e.g. fillers in SWT).
\item the developer choice to use a lightweight hack rather than a heavyweight strategy (e.g. empty labels versus complex layout objects in SWT).
\item the nature of the S-score that is sensible to the tyranny of the majority by construction.
\end{itemize}
To sum up, the high S-Scores that are due the tyranny of the majority still give us insights on the analyzed software.

\subsection{Finding: Missing method calls are often grouped}

When analyzing the issues raised by our system, it turns out that the strangest type-usages are often instances of the same issue, i.e. are deviations to the same the standard usage.
For instance, in the Apache Derby database system, they are 2 type-usages with a S-score higher than 0.8. Both of them are related to the same issue (a missing call to ''getTransactionExecute()''in the context of ''executeConstantAction(Activation)''). 
This is logical property of our algorithm:
if a given type-usage has 5 exactly-similar type-usages and 100 almost-similar type-usages, all exactly-similar type-usages share the same set of almost-similars. Hence, all 5 exactly-similar type-usages have the same high S-score and the same predicted missing method calls.

\subsection{Finding: Missing method calls may reveal encapsulation breaking}
\label{encapsulation-breaking}
In three cases among the strangest type-usages, the system finds issues related to object encapsulation, a particular case of the law of Demeter \cite{Lienberherr1989}. While our system is not designed to find such violations, it turns out that these violations are also caught by the S-score. For instance, let us consider the following excerpt of \code{TrustCertificateDialog}:
\begin{verbatim}
certificateChainViewer = new TreeViewer(composite, SWT.BORDER);
certificateChainViewer.getTree().setLayout(layout);
certificateChainViewer.getTree().setLayoutData(data);
\end{verbatim}
This code sets two times a property of an internal object of the \code{TreeViewer}.
This is a a violation of the the encapsulation provided by a \code{TreeViewer}\footnote{see \url{https://bugs.eclipse.org/bugs/show_bug.cgi?id=296568}}. 
Our system detects them because these two internal objects are reflected in bytecode with two different type-usages containing a single method call each (resp. \code{setLayout} and \code{setLayoutData}).
However, the majority rule or \code{Tree} objects states that there is never a single call to \code{setLayout} or \code{setLayoutData}.

\subsection{Finding: Missing method calls may reveal dead code}

Some missing method calls found by our system are related to type-usages that are part of dead code.
The reason is that dead code is neither used nor maintained, hence doesn't honor the latest method call protocols.
Here are two concrete examples on Eclipse:
\begin{itemize}
\item In the class \code{BuildOrderPreferencePage}, the system misses the usual call to \code{setLayout} on a \code{Composite}. Actually, this type-usage (called \code{noteComposite}) is never used (line 225 of version \code{I20080903-R34patches}).

\item In the class \code{TableRenderingPreferencePage}, the system finds two different strange type-usages. We found out that: 1) this class is never used except in another class that is also never used and that was last edited on March 27, 2005 (4 years ago). These classes are dead code.
\end{itemize}

\medskip
\textbf{Recapitulation: }These qualitative case-studies shows that our approach to detecting missing method calls as violations of the majority rule is efficient to find different kinds of issues in object-oriented software.

\section{Related Work}
\label{related}

The idea of finding almost similar but not exactly identical code has been explored in depth in code clone research (see \cite{Roy2007} and e.g. \cite{Gabel2008a,Nguyen2009a}). However, there is a large gap to be filled between finding code clones and predicting real bugs with a low false positive rate. 
For instance, the specification of an algorithm that translates a set of code clones into a valid missing method call is much work. This is what we do in this paper, on top of a new definition of code similarity.
Also, our approach works on the software topology that we define in a similar way as Robillard et al. \cite{Robillard2008}. To a certain extent, our approach proposes a definition of a certain kind of anomaly in the software topology. Furthermore, the notion of ``inconsistency'' of the same author \cite{Robillard2007} concerns inconsistent software evolution and not bugs as we consider in this paper.

Engler et al. \cite{Engler2001} presented a generic approach to infer errors in system code as violations of implicit contracts.
Their approach is more general-purpose than ours in the sense that we only detect a special kind of problems: missing method calls.
The corresponding advantage is that our approach is automatic and does not require a template of deviant behavior and the implementation of one checker per template.
The same argument applies for FindBugs \cite{Hovemeyer2004}, which also addresses low-level bugs and is successful only if an error pattern can be formalized.

Other pattern-specific approaches to detecting deviant code include the one from Williams and Hollingsworth \cite{Williams2005}, who propose an automatic checking of return values of function calls. Also, Chang et al. \cite{Chang2007} target neglected tests in conditional statements.  Weimer and Necula \cite{Weimer2004} detect failures to release resources or to clean up properly along all paths.
Those approaches are not directly comparable to ours since they search for  different kinds of issues that are not directly related to missing method calls.

Another interesting approach from the OS research community is PR-Miner \cite{Li2005}. 
PR-Miner addresses missing procedure calls in system code and not API-specific bugs as we do at the scope of each type-usage.
Further, PR-Miner uses frequent item set mining, which is a NP-hard problem \cite{Yang2004}; on the contrary, the computation of the sets of exactly-similar and almost-similar type-usages is done in polynomial time ($O(N^2)$, where $N$ is the total number of type-usages).

There are several techniques for finding defects in programs based on the analysis of execution traces. For instance, Ernst et al \cite{Ernst2001}, Hangal and Lam \cite{Hangal2002}, and Csallner et al. \cite{Csallner2008} mine for undocumented invariants. Yang et al \cite{Yang2006} and Dallmeier et al. \cite{Dallmeier2005} mine traces for ordered sequences of functions.
Since our approach is based on the static analysis of source code, our approach requires less input data: it needs neither large traces of real usages nor comprehensive test suites, which are both difficult and costly to obtain.

There are a number of approaches to finding bugs using historical artifacts of software (e.g. two different versions of the same software package or the full revision control data) \cite{Livshits2005,zimmermann-tse-2005,Kim2009,Nguyen2010a}. An important difference with our approach is that it does not have such requirements on the input artifacts.
For instance,  Livshits et al. \cite{Livshits2005} extract common patterns from software revision histories. Hence, to be able to catch a defect, the repository must contain 1) a large number of occurrences of the same kind of bug and 2) a large number of corrections of these bugs. Our approach does not have these requirements, it is able to catch a strange type-usage even if this kind of strange  code has occurred only once in the whole software history.
However, it is an interesting future work direction to combine our approach with history-based bug detection techniques.

Wasylkowski et al. \cite{Wasylkowski2007} searched for \emph{locations in programs that deviate from normal object usage -- that is, defect candidates}. 
Their definition of object usage anomalies is also based on method calls, but in a more complex manner: they take into account method call ordering and object states. Also, a major conceptual difference is that they mine explicit protocols, while our approach relies on observations only in an agnostic manner.
To our opinion, our approach is conceptually simpler, which makes it easier to be implemented in static analysis tools such as Coverity \cite{Bessey2010}. 
 
Nguyen et al. \cite{Nguyen2009} introduced the concept of \emph{groum} to refer to graph-based representation of source code. Their paper contains a proposal of using \emph{groums} to detect anomalies.
Our approach is conceptually completely different: we do not use the same abstraction over code (we use type-usages while they use groums), and our definition of anomaly and the intuition behind is much different as well. Since their evaluation has a rather different scope than ours\footnote{Their main case study is on Fluid, which is a research prototype developed by the same authors in the context of another research project. On the contrary, we applied our approach on Eclipse, which has been developed by senior IBM programmers over several years, and a significant number of our anomalies were validated and accepted as code patches by the Eclipse developers (which are not all related to us).}, we can not conclude on whether \emph{groums} or type-usages are more appropriate to find missing method calls, or whether one or the other has a less ratio of false positives in certain contexts. A comparative, generalizable empirical evaluation on different datasets is future work.

Finally, none of these related papers leverage the idea of simulating likely bugs to extensively explore the prediction space of the approach and thus achieve a large-scale evaluation.

\section{Conclusion}
\label{conclusion}

In this paper, we have presented a system, called DMMC, to detect missing method calls in object-oriented software. Providing automated support to find and solve missing method calls is useful at all moments of the software lifetime, from development of new software, to maintenance of old and mature software.
In particular, developers can use DMMC in three scenarios:
at development time, to help him/her solve certain bugs related to missing method calls (see Section \ref{sandra} and \ref{helping-sandra});
at code review time, in batch mode, to find problematic, ``strange'' places in code (see Section \ref{case-studies});
at maintenance time, when a bug is reported that seems related to missing method calls (see Section \ref{mcm-in-eclipse-bug-repository}).
In the first and the last case, knowing beforehand that  a bug is caused by a missing method call is impossible. In the presence of a difficult bug, DMMC is one tool in the debugging toolbox: if DMMC yields a high S-Score, the developer has a good chance to have at least a diagnostic and concrete analysis data: the almost-similar type-usages and the missing method calls with their likelihood.

The evaluation of the system showed that: 
1) the system gives a majority of correct results;
2) the high confidence warnings produced by the system are related to real missing method calls in mature software;
3) missing method calls often reveal issues that are larger in scope including software aging, cloning, and violations of API best practices.

One area of future work is to apply the concept of \emph{almost-similarity} not only to method calls but to other parts of software. For instance, searching for \emph{almost-similar} traces could yield major improvements in the area of runtime defect detections. Also, searching for \emph{almost-similar} conditional statements is worth further investigation to improve the resilience of software with respect to incorrect input.

\section*{Acknowledgements}
We would like to gratefully thank Marcel Bruch for his participation in the early days of this work and Eric Bodden for his support in writing the static analysis software.

\bibliographystyle{acmtrans}
\bibliography{strings,entries,bugs}

\newpage

\appendix

\section{Replication Guidelines}
\label{replication}

The datasets are based on:
\begin{enumerate}
\item Eclipse Classic Distribution v3.5  (\url{http://archive.eclipse.org/eclipse/downloads/drops/R-3.5-200906111540/eclipse-SDK-3.5-linux-gtk.tar.gz})
\item Apache Tomcat v6.0.20 (\url{http://svn.apache.org/repos/asf/tomcat/tc6.0.x/tags/TOMCAT_6_0_20})
\item Apache Derby revision 10000811 (\url{svn://svn.apache.org/repos/asf/db/derby/code/trunk})
\end{enumerate}

A copy of the datasets, as well as the type-usage extraction software and the evaluation software is available at \url{http://www.monperrus.net/martin/dmmc}.

\end{document}